\documentclass[aps,prb,twocolumn,superscriptaddress,showpacs,floatfix]{revtex4}

\usepackage{graphicx,color,epsfig,rotate}
\usepackage{graphicx}% Include figure files

\begin{document}

\title {Dispersive magnetic excitations in the $S=1$
antiferromagnet Ba$_3$Mn$_2$O$_8$}

\author{M. B. Stone}
\affiliation{Neutron Scattering Science Division, Oak Ridge National
Laboratory, Oak Ridge, Tennessee 37831, USA}

\author{M. D. Lumsden}
\affiliation{Neutron Scattering Science Division, Oak Ridge National
Laboratory, Oak Ridge, Tennessee 37831, USA}

\author{Y. Qiu}
\affiliation{NIST Center for Neutron Research, Gaithersburg,
Maryland 20899, USA}
\affiliation{Department of Materials Science and Engineering, University of Maryland, College Park, Maryland 20742, USA}

\author{E. C. Samulon}
\affiliation{Department of Applied Physics and Geballe Laboratory
for Advanced Materials, Stanford University, California
94305, USA}

\author{C. D. Batista}
\affiliation{Theoretical Division, Los Alamos National Laboratory, Los Alamos, New Mexico, 87545 USA}

\author{I. R. Fisher}
\affiliation{Department of Applied Physics and Geballe Laboratory
for Advanced Materials, Stanford University, California
94305, USA}

\begin{abstract}
We present powder inelastic neutron scattering measurements of
the $S=1$ dimerized antiferromagnet Ba$_3$Mn$_2$O$_8$.  The
$T=1.4$~K~magnetic spectrum exhibits a spin-gap of
$\Delta \approx 1.0$~meV and a dispersive spectrum with a bandwidth of approximately $1.5$~meV.
Comparison to coupled dimer models
describe the dispersion and scattering intensity
accurately and determine the exchange constants in Ba$_3$Mn$_2$O$_8$.
The wave vector dependent scattering
intensity confirms the proposed $S=1$ dimer bond.
Temperature dependent measurements of the magnetic excitations indicate
the presence of both singlet-triplet and thermally activated
triplet-quintet excitations.
\end{abstract}

\pacs{75.10.Jm,  %(Quantized spin models)
      75.40.Gb,  %(Dynamic properties)
      75.30.Et   % Exchange and superexchange interactions
      }

\maketitle

\section{INTRODUCTION}

Low-dimensional and gapped quantum magnets based upon strongly
coupled spin pairs or dimers with weaker interdimer interactions
have become especially topical systems.  This is primarily
due to the relevance of experimentally accessible
quantum critical points
\cite{Hertzprb1976,sebastiannature2006,tlcucl3a,phccprb}.
For antiferromagnetic intradimer exchange, the ground state of
such systems is a product of singlets, but strong magnetic fields can
close the spin-gap to excited triplet states via Zeeman splitting of
the triplet \cite{tlcucl3b,zheludevprb2007,stonenjp}.  Such systems thus
provide an elegant realization of a lattice gas of hardcore bosons in which
the external magnetic field plays the role of the chemical potential and the
interdimer coupling determines both the kinetic and potential energy of the
delocalized triplets
\cite{affleck1990,giamarchi1999}.  Depending on the balance of these energy
scales the triplets will either crystallize or condense at low
temperatures\cite{ricescience2000} or, under the right set of conditions,
form a supersolid \cite{batista2007}.  Ba$_3$Mn$_2$O$_8$ is
a particularly promising candidate material for the
detailed study of magnetic field dependent quantum critical points.
Ba$_3$Mn$_2$O$_8$
has been identified as a $S=1$ dimerized antiferromagnet and the low-temperature
phase diagram has been examined using thermodynamic measurements
\cite{uchidajphys2001,Tsujiiprb2005,samulonpaper}.  However, a measure of the
dominant exchange constants has been notably absent.

%Figure 1 Structure Figure.
\begin{figure}[t]
\centering\includegraphics[scale=0.8,angle=-90]{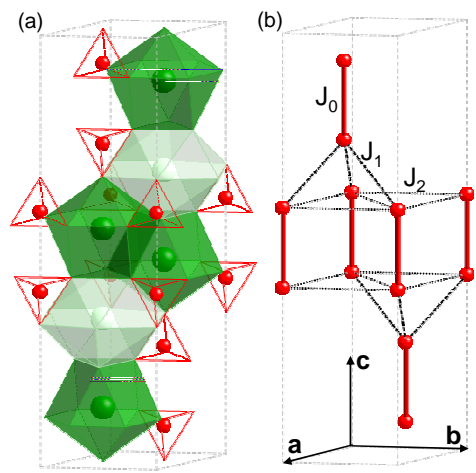}
\caption{\label{fig:structure}(Color online) Crystal structure
of Ba$_3$Mn$_2$O$_8$.  Oxygen atoms are not shown for clarity.
Grey dashed lines indicate chemical unit cell.
(a) Polyhedral representation of structure.
Light green and dark green polyhedra around Ba sites
represent the
two types of Ba coordination that alternate along the
$c$-axis, 12 and 10-fold coordination respectively.  Red tetrahedra
illustrate the Mn$^{5+}$ coordination.
(b) Simplified structure figure illustrating dominant exchange
connectivity in Ba$_3$Mn$_2$O$_8$ within and between the $S=1$ dimers.
$J_1$ and $J_2$ exchange correspond to distances of 4.569 and
5.711~\AA~respectively.}
\end{figure}

Low-field magnetic susceptibility and pulsed field magnetization
measurements of Ba$_3$Mn$_2$O$_8$ agree with weakly coupled
$S=1$ dimer models with antiferromagnetic dimer exchange
$J_0$ between 1.50 and 1.65~meV and a
zero-field spin-gap of $\Delta = 0.97$~meV
\cite{uchidajphys2001,uchidaprb2002}.  Unfortunately,
comparisons
to isolated dimer models
only yield results in terms of weighted sums of interdimer
exchange.  Specific heat measurements on powder
samples have been performed in
magnetic fields up to $\mu_0 H = 29$~T to describe the
low-temperature magnetic phase diagram of Ba$_3$Mn$_2$O$_8$\cite{Tsujiiprb2005}.  At
$T\approx 0.6$~K, a critical field of
$\mu_0 H \approx 11$~T is required to induce magnetic ordering.
Following the thermodynamic
phase transition to lower temperatures,
the experiments extrapolate to a zero-temperature
$\mu_0 H_{c1} = 9.04(0.15)$~T quantum critical point.
These measurements also indicate two phase transitions near
the lower critical field implying the existence of two
magnetic long-range-ordered phases in close proximity
to the quantum critical point.

More recent measurements using single crystals confirm the presence
of two distinct ordered phases and provide a detailed map of the
phase diagram \cite{samulonpaper}.  EPR measurements indicate a
zero-field splitting of the triplet states, attributed
in part to the effects of single ion anisotropy \cite{esrreff}, and
implying more complex magnetic structures than previously suggested\cite{uchidaprb2002}.

The applicable Hamiltonian for this system is not yet fully
characterized, and the magnitude and
extent of dimer and interdimer exchange interactions are
important for further understanding of Ba$_3$Mn$_2$O$_8$ in both
zero and applied magnetic fields.
Here we describe temperature
dependent inelastic neutron scattering (INS) measurements on
polycrystalline samples to determine the extent
of dispersive magnetic excitations and exchange
constants in Ba$_3$Mn$_2$O$_8$.

Ba$_3$Mn$_2$O$_8$ is hexagonal (space group $R\bar{3}m$)
with room temperature lattice constants $a=5.711$ and
$c=21.444$~\AA \cite{weller1999}.  The Mn$^{5+}$ ions
reside in a distorted tetrahedral environment, as shown
in Fig.~\ref{fig:structure}(a), resulting in an effective $S=1$ moment.
Dimer and interdimer magnetic interactions are considered to be
antiferromagnetic ($J>0$) and
Heisenberg with a small single-ion anisotropy, $D$.
Including a Zeeman term for applied magnetic fields $H$ along the
$z$-axis, the Hamiltonian is
\begin{equation}
\label{eq:hamiltonian}
\mathcal{H} = \sum_{i,j}\frac{J_{i,j}}{2}\mathbf{S}_i \cdot \mathbf{S}_j
+ D\sum_i (S_i^z)^2 - g\mu_B H \sum_i S_i^z,
\end{equation}
where $i$ and $j$ designate coordinates of individual interacting
spins, $S$.
The nearest neighbor $S=1$ sites along the $c$-axis, $d = 3.985$~\AA,
have been proposed as strongly coupled antiferromagnetic $S=1$ dimers.
The dimers form an edge-sharing triangular lattice bilayer
in the $ab$-plane.
Neighboring planes along the $c$-axis are separated by alternating
layers of oxygen coordinated Ba sites, ten and twelve site coordination
respectively, as illustrated in Fig.~\ref{fig:structure}(a).  Weaker
interdimer exchange within the $ac$-plane, $J_2$, and between the
bilayers, $J_1$, have also been proposed based upon the crystal structure.  These are illustrated in Fig.~\ref{fig:structure}(b) along
with the dimer exchange $J_0$.

\section{EXPERIMENTAL METHODS}

Powder samples of Ba$_3$Mn$_2$O$_8$ were synthesized
using stoichiometric amounts of
BaCO$_3$ and Mn$_2$O$_3$ in a solid state reaction.
The reactants were calcined under flowing oxygen at
$900$~$^\circ$C~for 30~h.  The resulting green powder was
then reground and sintered between $900$ and $1000$~$^\circ$C
under flowing oxygen for approximately 7 days.  This growth
procedure is similar to that described in
Ref.~\onlinecite{uchidajphys2001}.
Single crystals of appropriate mass are unfortunately not yet
available for studies of the dispersion using INS.

SQUID magnetization measurements as a function of temperature for
$\mu_0 H = 1000$~Oe~and as a function of magnetic field for $T=40$~K~
did not reveal any measurable impurities due to Mn$_2$O$_3$ or
Mn$_3$O$_4$ which are ferrimagnetic below $T_c = 79$~K and antiferromagnetic below $T_c = 43$~K respectively\cite{robie1985}.
The temperature dependent magnetic susceptibility compares well
with previously published data with a
rounded peak at $T=18$~K~and an
activated low-temperature susceptibility characteristic of
antiferromagnetic spin-gap systems \cite{uchidajphys2001,uchidaprb2002}.
We also characterize the $150 \leq T \leq 350$~K~magnetic
susceptibility via a Curie-Weiss law with
$\Theta_{\mathrm{CW}}=-43.2(2)$~K.

Inelastic neutron scattering measurements were performed on 76
grams of Ba$_3$Mn$_2$O$_8$ in an 18 mm diameter and 100 mm tall
cylindrical aluminum sample
can.  The sample was produced from five separately prepared batches of
Ba$_3$Mn$_2$O$_8$.  Each batch was checked for impurity phases using SQUID
magnetization and powder X-ray diffraction.  Both measurements found no
measurable impurity phases.
Inelastic neutron scattering measurements were performed using the direct geometry
chopper spectrometer, DCS, at the NIST Center
for Neutron Scattering.  Spectra
were measured for temperatures between $T=1.4$ and $T=160$~K for one
hour in each configuration.  Temperature control was provided by
a liquid He$^4$ flow cryostat.  Two incident wavelengths, $\lambda$, were used.
$\lambda = 2.9$~\AA~measurements probed energy transfers up to $\hbar\omega = 5.5$~meV~and wave vectors
up to $Q = 4.05$~\AA$^{-1}$~at the elastic position.
$\lambda = 4.4$~\AA~measurements provide improved energy and wave vector
resolution up to $\hbar\omega = 3.05$~meV and $Q = 2.69$~\AA$^{-1}$.  The calculated full width at half maximum (FWHM) energy
resolution at the elastic position
is $\delta \hbar \omega = 0.5$ and $0.15$~$\mu$eV~for
the $2.9$ and $4.4$~\AA~incident wavelengths respectively.
Background measurements were made for each incident wavelength using an empty sample can at $T=1.4$~K.  A $T<200$~K~vanadium standard was
measured for calibration of detector sensitivity.  A scattering
angle dependent absorption correction for the cylindrical sample
geometry was also applied to the inelastic scattering intensity.
Unless otherwise noted, these backgrounds and normalization are applied to our presented results.

Neglecting Bose occupation and Debye-Waller factors, the magnetic neutron-scattering cross section is proportional to
the scattering function, $\mathcal{S}(\mathbf{Q},\omega)$,
\begin{equation}
\label{eq:crosssection}
\frac{d^2 \sigma}{d\Omega dE^{\prime}} \propto \frac{k^{\prime}}{k}|F(Q)|^2 \mathcal{S}(\mathbf{Q},\omega),
\end{equation}
where $k^{\prime}$ and $k$ are the magnitude of the final and initial
neutron wave vectors and $F(Q)$ is the magnetic ion form factor.  We
plot our measured scattering intensity in units of $\mathcal{S}(\mathbf{Q},\omega)$, but we
do not normalize the data by the magnetic form factor.

\section{RESULTS AND DISCUSSION}

\begin{figure}[t]
\centering\includegraphics[scale=0.875]{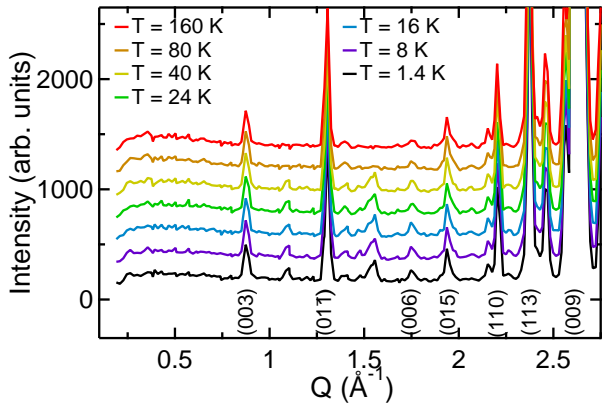}
\caption{\label{fig:diffractdata}
(color online) Temperature dependent diffraction for Ba$_3$Mn$_2$O$_8$
measured in the $\lambda=2.9$~\AA~configuration.  Data are integrated
between -0.1 and 0.1 meV energy transfer.  Data have not
been background subtracted or corrected for absorption.  Higher
temperature data are offset vertically from the $T=1.4$~K data for presentation.  Several characteristic Bragg peaks
are indexed in the figure.}
\end{figure}

Figure~\ref{fig:diffractdata} shows the scattering intensity as a
function of wave vector transfer in the vicinity of the elastic
position, $-0.1 \leq \hbar \omega \leq 0.1$~meV, at several
temperatures.  Nuclear Bragg peak positions and intensity compare
well with the previously determined room-temperature structure
\cite{weller1999}.  We note that for $T\leq 80$~K~there are two
additional weak Bragg peaks at $Q=1.096$ and $1.545$~\AA$^{-1}$.
These are likely associated with either Mn$_2$O$_3$ or Mn$_3$O$_4$ impurities as discussed earlier.  We do not observe any contribution from spin-waves due to these impurities in the inelastic portion of the spectra.

%Figure2 Diffraction pattern vs. temperature

%Figure3 lattice parameters vs. temperature
\begin{figure}[t]
\centering\includegraphics[scale=0.875]{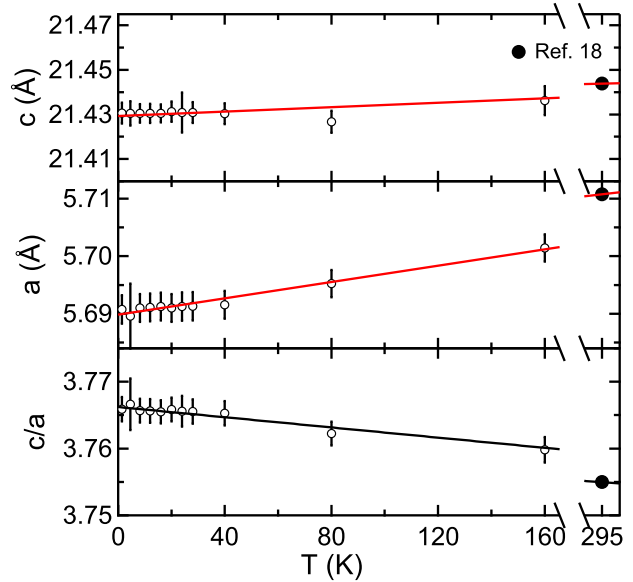}
\caption{\label{fig:latticeparam}
(color online) Temperature dependent lattice parameters of Ba$_3$Mn$_2$O$_8$.
(a) $c$-axis lattice constant versus temperature.  (b) $a$-axis
lattice constant versus temperature.  (c) $c/a$ as a function of
temperature.  Solid points from Ref.~\onlinecite{weller1999}
for $T=295$ K are shown for comparison.  Lattice constants were obtained
from the weighted mean peak position of the (003), (006), (009) and
(110) Bragg peaks for $\lambda=2.9$ and $4.4$~\AA.  Error bars are the error in the mean.  Solid lines in (a) and (b)
are linear fits described in the text.  Solid line in (c) is
calculated based upon the linear fits for the $a$ and $c$-axis
lattice constant.}
\end{figure}

Although wave vector resolution was not optimized for diffraction
($\frac{\delta Q}{Q=1 \mathrm{\AA}} \approx 0.03$ for $\lambda = 4.4$~\AA), we
fit the $(003)$, $(006)$, $(009)$ and $(110)$ Bragg peak positions for
each incident wavelength to determine lattice constants as a
function of temperature, \textit{c.f.} Fig.~\ref{fig:latticeparam}(a)
and (b).  These values are consistent with the previously determined
room temperature structure.  There are no apparent structural phase transitions from $T=160$~to 1.4~K.  Both the
$a$ and $c$ lattice vectors contract at lower temperatures, but
there is only a $0.3\%$ change in the $a$-axis lattice constant and
an even smaller change in the $c$-axis lattice constant, $0.06\%$,
from $T=295$~K~to $T=1.4$~K.  Linear fits to these data along with the $T=295$~K~values are shown in Fig.~\ref{fig:latticeparam}(a) and (b) along
with calculated values of $c/a$ and the respective calculated
curve in panel (c).  The fitted lines
provide a good description of the data
with coefficients of linear expansion:
$\alpha_c = \frac{1}{c(T=0)}\frac{dc}{dT} = 2.3(1)E-6$ and
$\alpha_a = \frac{1}{a(T=0)}\frac{da}{dT} = 1.25(1)E-5$~K$^{-1}$.  Including
quadratic terms does not substantially improve the fits.  The weak lattice parameter temperature dependence indicates
that any changes in the magnetic excitation spectra with
temperature are not likely associated with changes in exchange
due to changes in distance between individual spins.

%Figure 4 Powder Spectrum Figure.
\begin{figure*}[t]
\centering\includegraphics[scale=0.90]{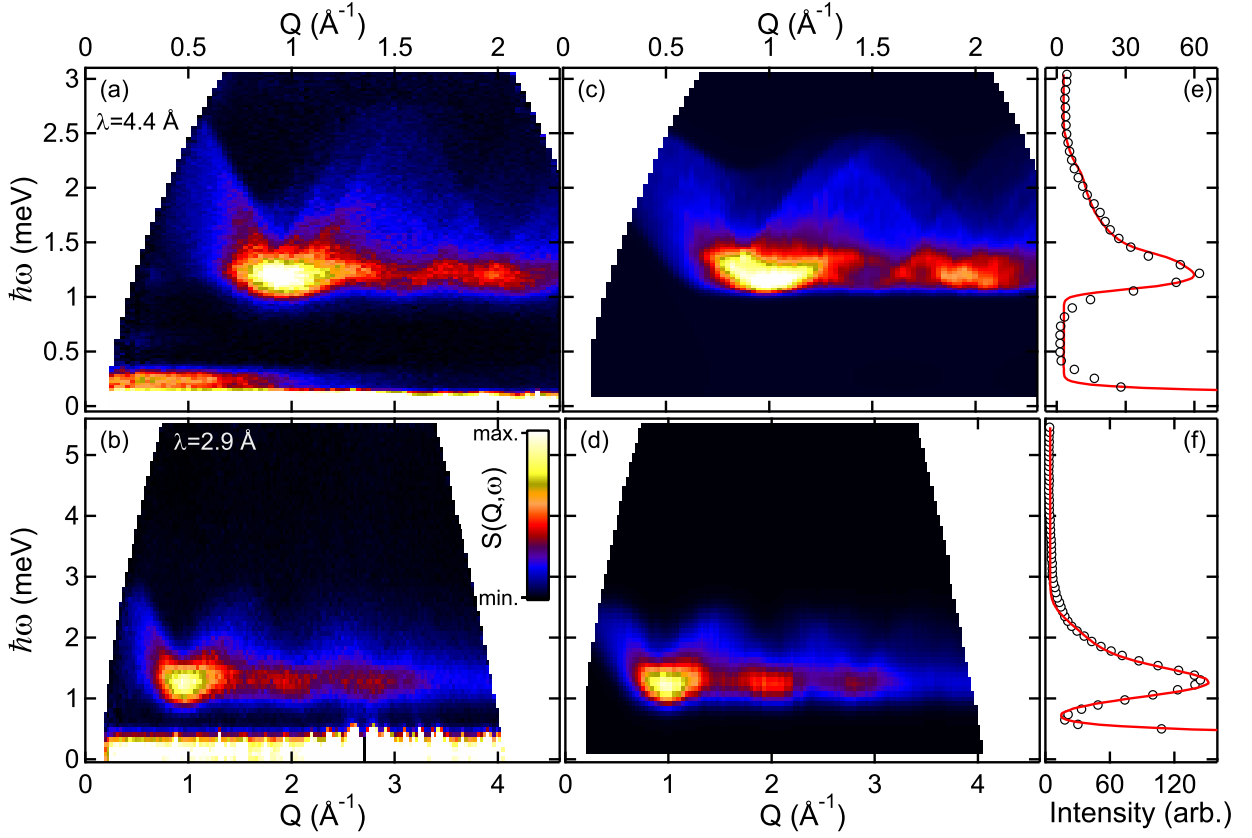}
\caption{\label{fig:contourfig}(Color online) Measured and
calculated INS intensity for
Ba$_3$Mn$_2$O$_8$ at $T=1.4$~K.  Panels (a) and (b) are data measured
for incident wavelengths of $\lambda = 4.4$ and $2.9$~\AA~respectively.  Panels (c) and (d) are the calculated
scattering intensities corresponding to the
coupled $S=1$ dimer model described in the text.  Panels (e) and (f) are the energy dependent integrated scattering intensity of the data and model calculations shown in panels (a)-(d) for wave vectors $0.65 < Q < 2$~\AA$^{-1}$ and $0.75 < Q < 3.25$~\AA$^{-1}$ for $\lambda=4.4$ and $2.9$~\AA~respectively. Model lineshapes in (e) and (f) are fit to
the data including an elastic Gaussian lineshape and a constant background as discussed in the text.}
\end{figure*}

The $T=1.4$~K~excitation spectra for the $\lambda=4.4$ and
$2.9$~\AA~configurations are shown in Fig.~\ref{fig:contourfig}(a)
and (b).  There is a single dispersive excitation
with a gap of $\Delta \approx 1$~meV and a bandwidth of
approximately 1.5~meV.  The decreasing
scattering intensity with increasing wave vector
transfer immediately
suggests the spectrum is magnetic.  The increase
in scattering intensity near $\hbar\omega=0.25$ meV in
Fig.~\ref{fig:contourfig}(a) for small wave vectors is not
intrinsic to the sample and
has been
seen in background measurements.
Figure~\ref{fig:contourfig}(e) and (f) show the low-temperature
wave vector integrated scattering intensity.  There is a peak
in the magnetic density of states in the vicinity of the
spin-gap energy, but there is no peak at the top of the
dispersive band of excitations.  The spectrum of one-dimensional
Heisenberg gapped quantum antiferromagnets has
characteristic Van-Hove singularities at
the top and bottom of the band.  The absence of
any peak at higher energy transfers implies the magnetic excitations
in Ba$_3$Mn$_2$O$_8$ are at least two-dimensional in
their connectivity.

For comparison to our results, we calculate the INS cross section of
$S=1$ antiferromagnetic dimers with weak interdimer interactions.
At finite temperature the scattering function is written as
\cite{squires}
\begin{eqnarray}
\label{eq:sqwfull}
\mathcal{S}(\mathbf{Q},\omega) &=&
\frac{1}{Z}\sum_{\psi , \psi^{\prime}}e^{- E_\psi \beta} \sum_j |\langle\psi^{\prime}|S_j e^{i \mathbf{Q} x_j}|
\psi\rangle|^{2} \nonumber \\
& & \delta(E_{\psi} - E_{\psi^{\prime}}+
\hbar\omega),
\end{eqnarray}
where $\beta = \frac{1}{k_B T}$, $Z$ is the partition function
and the sum is over the initial
and final states $\psi$ and $\psi^{\prime}$ with
energies $E_{\psi}$ and $E_{\psi^{\prime}}$.  For a dimer, Eq.~\ref{eq:sqwfull} becomes
\begin{eqnarray}
\label{eq:sqwsum}
\mathcal{S}(\mathbf{Q},\omega) &=& \frac{1}{Z}\sum_{\psi ,
\psi^{\prime}}e^{- E_\psi \beta}
|\langle\psi^{\prime}|S_1 e^{i \mathbf{Q} x_1} + S_2 e^{i \mathbf{Q} x_2}|
\psi\rangle|^{2} \nonumber \\
& & \delta(E_{\psi} - E_{\psi^{\prime}}+
\hbar\omega),
\end{eqnarray}
where $x_1$ and $x_2$ are the respective
crystallographic coordinates of the spins
in the dimer and $S_n$ are the spin operators.

An isolated antiferromagnetic $S=1$ dimer with intradimer exchange $J_0$
will have a non-magnetic (total spin $S_T = 0$) singlet ground state
at an energy of $E=-2J_0$ with
$S_T = 1$ triplet and $S_T = 2$ quintet excited states at energies
$E=-J_0$
and $E=J_0$.  This results in triplet and quintet spin-gaps of
$\hbar \omega=J_0$ and $\hbar \omega = 3 J_0$.  However, magnetic INS will only
probe the singlet-triplet and triplet-quintet cross-sections, \textit{i.e.} $|\Delta S_{T}|=1$.  Including structure factors from
Eq.~\ref{eq:sqwsum}, the scattering function is
a sum of two terms,
\begin{eqnarray}
\label{eq:weightedsqw}
\mathcal{S}(\mathbf{Q},\omega) &=&
\frac{4e^{2J\beta}[1-\cos({\mathbf{Q}\cdot \mathbf{d}})]}{e^{2J\beta} + 3e^{J\beta} + 5e^{-J\beta}}\delta(\hbar\omega - J) + \\ \nonumber
& &
\frac{5e^{J\beta}[1-\cos(\mathbf{Q} \cdot \mathbf{d})]}{e^{2J\beta} + 3e^{J\beta} + e^{-J\beta}}\delta(\hbar\omega - 3J),
\end{eqnarray}
where $\mathbf{d}$ is the bond vector between the spins of the dimer.
Although the calculated matrix elements of the triplet-quintet
transitions are larger than the singlet-triplet transitions,
the triplet-quintet transitions are thermally
activated with less spectral weight and will only be populated at
higher temperatures.

Equation~\ref{eq:weightedsqw} does not account for interdimer
correlations,  \textit{i.e.} dispersive excitations.
The random phase
approximation (RPA) has been successful in describing
the dispersion of weakly coupled dimers in spin-gap
systems.  This has been illustrated for varying numbers
of interactions and spin-quanta in several experimental
systems including KCuCl$_3$\cite{katojphyssocjpn1998,cavadiniEphysJB},
TlCuCl$_3$\cite{cavadiniPRB2001}, Cs$_3$Cr$_2$Br$_9$
\cite{leuenbergerprb1984}, Cs$_3$Cr$_2$I$_9$\cite{leuenbergerprb1986},
PHCC\cite{phccprb} and BaCuSi$_2$O$_6$\cite{sasagoprb1997}.
The RPA dispersion for Heisenberg exchange coupled dimers is
\begin{equation}
\label{eq:rpadisp}
\hbar\omega(\mathbf{Q}) = \sqrt{\Delta^{2} + M^2\Delta\mathcal{J}(\mathbf{Q})R(T)}
\end{equation}
where $M^2$ is the transition matrix element ($M^2 = \frac{4}{3}S[S+1]$),
$\mathcal{J}(\mathbf{Q})$ is the Fourier sum over interactions beyond
dimer exchange, $\Delta \equiv J_0$ and $R(T)$ is
the thermal population difference between the ground and excited states.  For $S=1$ antiferromagnetic dimers,
\begin{equation}
\label{eq:roftdisp}
R(T) = \frac{1-\exp(-\Delta\beta)}{1+3\exp(-\Delta\beta) + 5\exp(-\Delta\beta)},
\end{equation}
considering only singlet-triplet excitations.
Ba$_3$Mn$_2$O$_8$ has a single dimer per unit cell and the interdimer interactions
propagate the triplet excitation leading to the Fourier sum
\begin{eqnarray}
\label{eq:dispeqn}
\nonumber \omega_2 &=& \cos(2\pi k) + \cos(2\pi[h+k]) + \cos(2\pi h) \\
\nonumber \omega_1 &=& \cos(\frac{2\pi}{3}[-h+k+l]) + \cos(\frac{2\pi}{3}[-h-2k+l]) \\
\nonumber & &
+ \cos(\frac{2\pi}{3}[2h + k +l])  \\
\mathcal{J}(\mathbf{Q}) &=& 2 J_2\omega_2 + J_1\omega_{1}.
\end{eqnarray}
Recent EPR measurements have revealed a zero-field splitting of
$D =-0.032$~meV, although both modes will have an identical
dispersion\cite{esrreff}.  Examination of
thermodynamic measurements have included
an exchange constant, $J_3$, which couples neighboring dimers
in a bilayer from spin-1 of a dimer to spin-2 of a second
dimer\cite{uchidaprb2002}.  This exchange term represents a change in phase of the
triplet excitation between dimers in the $ab$ plane, and would change the prefactor
of the $\omega_2$ term in Eq.~\ref{eq:dispeqn} to be $2(J_2 - J_3)$.
We have chosen to omit the $J_3$ exchange from the current analysis.
Because of its large spin-spin distance ($6.964$~\AA) and out of plane
coupling, it is presumably much weaker than $J_2$
and $J_1$. In addition, recent calculations examining the relative strength of
exchange constants in Ba$_3$Mn$_2$O$_8$ have shown that the $J_3$ exchange
constant is effectively zero \cite{koo2006}.

The scattering function must also be modified to account for the
dispersive excitations.  The single mode approximation (SMA) has been successfully applied to
several dispersive disordered gapped antiferromagnets
with interdimer
exchange included in the Hamiltonian
\cite{hohenbergbrinkman,phccprb,maprl1992,sasagoprb1997,taohongprb}.
This results in an additional multiplicative
$\frac{1}{\omega(\mathbf{Q})}$ term in $\mathcal{S}(\mathbf{Q},\omega)$
such that considering only singlet-triplet scattering for
Ba$_3$Mn$_2$O$_8$ the scattering function becomes
\begin{equation}
\label{eq:fullsofqwsma}
\mathcal{S}(\mathbf{Q},\omega) =
\frac{4e^{2J\beta}[1-\cos({\mathbf{Q}\cdot \mathbf{d}})]}{(e^{2J\beta} + 3e^{J\beta} + 5e^{-J\beta})\hbar\omega(\mathbf{Q})}\delta(\hbar\omega - \hbar\omega(\mathbf{Q})),
\end{equation}
where $\hbar\omega(\mathbf{Q})$ is given by Eqs.~\ref{eq:rpadisp}-\ref{eq:dispeqn}.
The SMA is appropriate for the case of Ba$_3$Mn$_2$O$_8$
for $T\ll J_0$ since only singlet-triplet excitations will be
effectively probed via INS in this temperature regime.

For comparison of our high resolution ($\lambda = 4.4$~\AA)
polycrystalline measurements we
numerically \emph{spherically average} Eq.~\ref{eq:fullsofqwsma},
\begin{equation}
\mathcal{S}(Q,\omega) = \int\frac{d\Omega_{\hat{q}}}{4 \pi}\mathcal{S}(\mathbf{Q},\omega).
\end{equation}
This process was also recently employed successfully in examination of
polycrstyalline measurements of a gapped antiferromagnet
using the same instrumentation\cite{taohongprb}.  The interpretation
of our measurements is more straightforward given the absence of
hydrogen or contamination from phonons in our spectrum.  We calculate $\mathcal{S}(\mathbf{Q},\omega)$ over
spherical shells in $|\mathbf{Q}|$ space at
fixed values of energy transfer
with $\mathbf{d}$ fixed as the proposed dimer bond vector.
This was done for a series of $J_0$, $J_1$ and
$J_2$ values.  This spectrum was combined with an identical spectrum shifted in
energy transfer by
the value $D=-0.032$~meV.  The spectrum was then
multiplied by $|F(Q)|^2$~~\cite{magformfacnote,pjbrown} and convolved with a Gaussian
representation of the mean instrumental energy and wave vector resolution
over the energy and wave vector range of the magnetic excitation.
A constant background and multiplicative prefactor were
used as fitting parameters of the calculated spectrum in comparison
to the measured data.  This procedure yields best fit exchange
constants
$J_0 = 1.61(3)$,~$J_1 = -0.062^{+0.007}_{-0.066}$~
and~$J_2 = 0.112^{+0.015}_{-0.003}$~meV\cite{errorbar}.  The
corresponding
best fit INS scattering intensity for both instrument configurations
is plotted in Fig.~\ref{fig:contourfig}(c) and (d),
and agrees very well with the dispersion and intensity modulation
observed in the measurement.  The determined value of $J_0$
and the corresponding spin-gap based upon the dispersion,
$\Delta = 1.05$~meV, are both in the vicinity of values
from thermodynamic measurements.

Figures~\ref{fig:contourfig}(e) and (f)
show the fitted wave vector integrated lineshapes compared
to the measured data.  Fit parameters for the calculated lineshapes
include an elastic Gaussian,
an overall multiplicative prefactor and a constant background.  The calculations
based upon both incident wave lengths agree very well with
the measurement.

\begin{figure}[t]
\centering\includegraphics[scale=0.80]{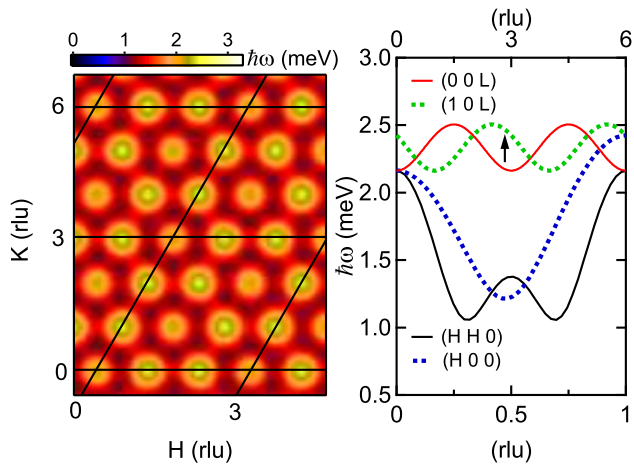}
\caption{\label{fig:crystdispfig}(Color online) $T=1.4$~K
singlet-triplet single crystal dispersion relation determined
for Ba$_3$Mn$_2$O$_8$ based upon powder average analysis.
(a) Contour figure of dispersion in the hexagonal (HK0)
plane.
(b) Dispersion as a function of reciprocal
lattice units (rlu) along other primary directions
of the hexagonal structure.  Dispersion along
the $(0~0~L)$ and $(1~0~L)$ directions is plotted using the
top axis.}
\end{figure}

We plot the low-temperature singlet-triplet
dispersion relation in Fig.~\ref{fig:crystdispfig}.
Overall minima in the dispersion occur at the
$(\frac{1}{3}-\delta,~\frac{1}{3}-\delta,~l)$ and
$(\frac{2}{3}+\delta,~\frac{2}{3}+\delta,~l)$ wave vectors for $l=3n$ and
$(\frac{1}{3}+\delta,~\frac{1}{3}+\delta,~l)$ and
$(\frac{2}{3}-\delta,~\frac{2}{3}-\delta,~l)$ wave vectors for
$l=\frac{3}{2} + 3n$ where $n$ is an integer and
$\delta \approx 0.025$.
The nonzero value
of $J_1$ results in  a finite  dispersion
along the $(\zeta~\zeta~l)$ direction with
a periodicity of three reciprocal
lattice units (rlu).  This is
shown in Fig.~\ref{fig:crystdispfig}(b) for the
curves plotted using the top axis.

%Figure 3 Single Crystal Dispserion.

%Figure 3 First Moment Figure.
\begin{figure}[t]
\centering\includegraphics[scale=0.90]{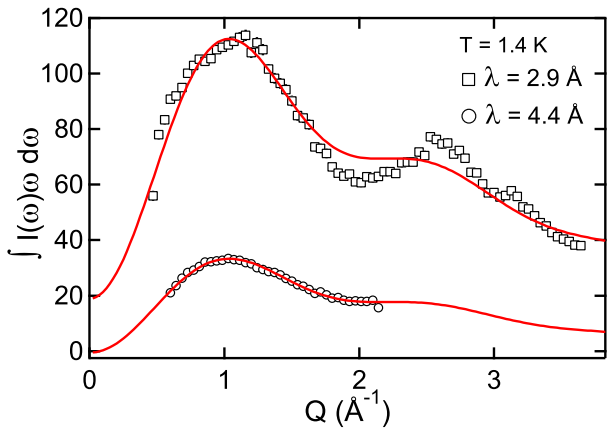}
\caption{\label{fig:fmfig}(Color online) First moment
of scattering intensity integrated between $\hbar\omega=0.75$
and $3$ meV for the $T=1.4$ K $\lambda=4.4$ and 2.9 \AA~measurements
shown in Fig.~\ref{fig:contourfig}(a) and (b).  Solid lines
correspond to fits described in the text with the fitted $|\mathbf{d}|$
value of $d= 4.073(7)$~\AA.}
\end{figure}

The powder average of the first frequency moment of the
$T=0$ energy integrated scattering function,
\begin{eqnarray}
\label{eq:firstmom}
\nonumber \hbar\langle \omega \rangle_Q  &\equiv& \int_{-\infty}^{\infty} \int \frac{d\Omega d\omega}{4\pi}  \hbar\omega(\mathbf{Q}) \mathcal{S}(\mathbf{Q},\hbar\omega)  \\
& \propto & |F(Q)|^2 [1 - \frac{\sin(Qd)}{Qd}],
\end{eqnarray}
provides direct information regarding the
length of the dimer bond, $d$.  Figure~\ref{fig:fmfig} shows
the first moment of the measured $T=1.4$~K~scattering intensity
for both incident wavelengths.  These data are
fit to Eq.~\ref{eq:firstmom} including an overall constant and
multiplicative prefactor.  A simultaneous fit of both data yields a
good representation of the measured results with
$d=4.073(7)$~\AA.  This value agrees
with the description of the
dimer bond being the short vertical bond between
Mn$^{5+}$ moments in the Ba$_3$Mn$_2$O$_8$
crystal structure $d = 3.985$ \AA,~\emph{c.f.}~Fig.~\ref{fig:structure}.

%Figure 4 Temperature dependent Scattering intensity.
\begin{figure}[t]
\centering\includegraphics[scale=0.95]{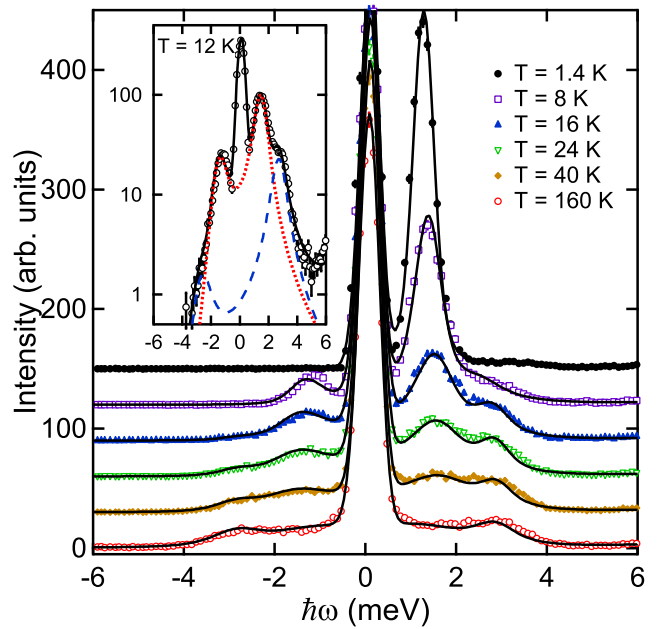}
\caption{\label{fig:inttempdepend}(Color online) Energy dependent
scattering intensity for $0.9 < Q < 1.0$~\AA$^{-1}$~for a series
of temperatures as measured in the $\lambda = 2.9$~\AA~configuration.  Data for different temperatures are vertically offset.  Solid lines are fits to the
Lorentzian functions described convolved with the energy
dependent resolution of the instrument.
Inset illustrates scattering intensity of the $T=12$~K measurement
plotted on a logarithmic intensity scale.  Dotted red line corresponds to
the spectral contribution of the lower energy mode
and the dashed blue line corresponds to the higher energy mode.}
\end{figure}

The magnetic spectra of Ba$_3$Mn$_2$O$_8$ as a function
of temperature for $1.4 < T \leq 160$~K was
also measured.
In order to consider a range of energy transfers
beyond the singlet-triplet band,
we examine the $\lambda=2.9$~\AA~ configuration data.
Constant
wave vector scans for $0.9 \leq Q \leq 1.0$~\AA$^{-1}$~are
shown in Fig.~\ref{fig:inttempdepend} for several temperatures
for both neutron energy gain and loss.  This
wave vector was chosen because it is in the vicinity
of both the peak in the density of states and the overall
minimum in the measured dispersion.  From the $T=1.4$~K~
spectrum in Fig.~\ref{fig:contourfig}(b),
this wave vector also
provides minimal interference from higher energy portions
of the singlet-triplet spectrum.
The $T=1.4$~K~spectrum consists of a single peak
at $\hbar\omega \approx 1.25$~meV~with no significant
scattering intensity on the neutron energy gain side of
the spectrum.  As temperature is increased, the single
peak broadens, moves
to slightly larger energy transfers and additional
scattering intensity develops in the vicinity of 3 meV.
There is also additional scattering intensity which
develops on the neutron energy gain side of the spectrum at
elevated temperatures.

In order to distinguish singlet-triplet
and thermally activated triplet-quintet excitations, we model
the constant wave vector scans using
two excitations.  We fit to two
Lorentzian
functions of the form \cite{zaliznyakprb1994}
\begin{eqnarray}
\label{eq:tempdependint}
\nonumber I(\omega) &=& \langle n(\omega)+1\rangle
(\frac{A\Gamma_1}{(\omega-\omega_1)^2 + \Gamma_1^2} - \frac{A\Gamma_1}{(\omega+\omega_1)^2 + \Gamma_1^2} \\
 & +& \frac{B\Gamma_2}{(\omega-\omega_2)^2 + \Gamma_2^2} - \frac{B\Gamma_2}{(\omega+\omega_2)^2 + \Gamma_2^2}),
\end{eqnarray}
where $\Gamma_1$, $\omega_1$, $\Gamma_2$ and $\omega_2$
are the half width at half maximum and energy of two respective
excitations with the Lorentzian width
reflecting temperature dependent broadening of the spectrum.
The Bose factor
$\langle n(\omega)+1\rangle \equiv [1-\exp(-\hbar\omega\beta)]^{-1}$
enforces detailed balance of the scattering intensity.  Equation~\ref{eq:tempdependint} and a variable width and
amplitude Gaussian peak at
the elastic position
were convolved with the energy transfer dependent instrumental
energy resolution with fitting parameters $A$, $B$,
$\Gamma_{1,2}$ and $\omega_{1,2}$.  A time independent background
of the form  $\mathrm{BG} \propto (\frac{81.81}{\lambda^2}-\hbar\omega)^{-2}$, as typically used
for time-of-flight direct geometry INS measurements and a
constant background were determined from the $T\leq 40$~K~data
and held fixed.  The two excitation fits
are shown as solid
lines in Fig.~\ref{fig:inttempdepend}.  The lineshapes agree
well with the energy dependent distribution of scattering intensity for
both neutron energy loss and gain.
To further illustrate the presence of the triplet-quintet excitation, we plot the $T=12$~K data and fitted lineshape including the individual contributions from
each mode in the inset of Fig.~\ref{fig:inttempdepend}.

%Figure 5 Temp dependent parameteres.
\begin{figure}
\centering\includegraphics[scale=0.95]{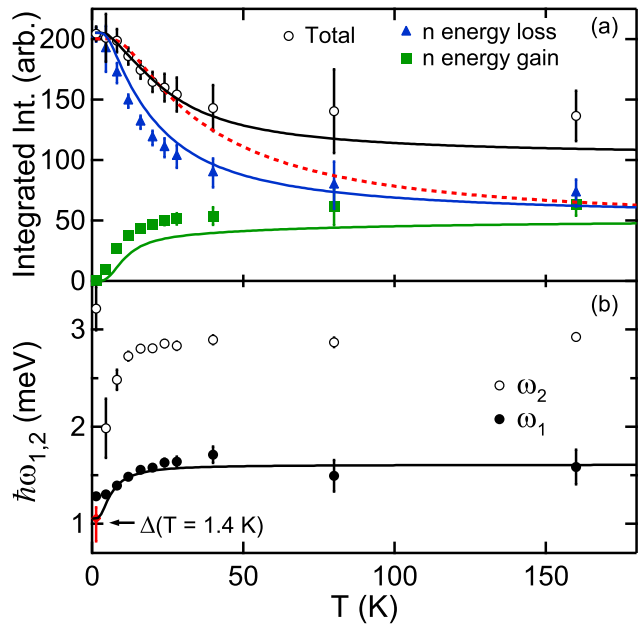}
\caption{\label{fig:paramtempdepend}(Color online) Temperature dependent
integrated intensity and characteristic energy
determined from fits to data shown in Fig.~\ref{fig:inttempdepend}.
(a) Total and neutron energy gain and loss portions of the integrated
scattering intensity for both modes.  The solid
black line is a fit of the temperature dependent total scattering
intensity of the isolated antiferromagnetic $S=1$ dimer including
both singlet-triplet and thermally activated triplet-quintet
excitations.  The dashed line only includes singlet-triplet excitations.
Solid blue and green curves are calculated from the singlet-triplet and
triplet-quintet fitted scattering intensity.  (b)
Mode energy as a function of temperature.  The $T=1.4$ K spin-gap value
determined from fitted exchange parameters is shown as a single
red diamond.  Solid line is a calculation of the RPA model temperature
dependent spin-gap using the determined values of exchange constants.}
\end{figure}

Figure~\ref{fig:paramtempdepend}(a) shows the total
integrated scattering intensity for both modes
as well as the neutron energy
gain and loss portions as a function of temperature.
We fit the total scattering intensity to the temperature dependent
isolated dimer scattering function in Eq.~\ref{eq:weightedsqw} with an overall multiplicative prefactor and the exchange $J_0$ as fitting
parameters.  Considering only singlet-triplet excitations yields the
dashed red line in Fig.~\ref{fig:paramtempdepend}(a) and a value of
$J_0 = 3.9(2)$ meV.  However, including both singlet-triplet and
triplet-quintet modes yields a much better fit (solid black line)
and a more appropriate value of $J_0 = 1.9(3)$ meV.  The improved fit
implies that there is a non-negligible contribution of thermally
activated triplet-quintet excitations at higher energy
transfers.  The calculated
temperature dependent scattering intensity for the
neutron energy
gain and loss portions of the spectrum based upon the
singlet-triplet, triplet-quintet model also agrees well the
data shown in Fig.~\ref{fig:paramtempdepend}(a).

The fitted values $\omega_1$~and~$\omega_2$~are
plotted in Fig.~\ref{fig:paramtempdepend}(b).  The $\omega_1$ value is not the spin-gap energy, rather it is the characteristic energy
of the lower energy mode found for the constant wave vector scan.
For comparison, we also plot the determined spin-gap energy based upon
the $T=1.4$~K~powder average analysis.  The calculated temperature dependent spin-gap of the singlet-triplet excitation based upon
Eqs.~\ref{eq:rpadisp}-\ref{eq:dispeqn} and the fitted exchange constants
is plotted as a solid line
in Fig.~\ref{fig:paramtempdepend}(b).  The RPA coupled dimer
description agrees well with the temperature dependence of the energy scale of the singlet-triplet excitation.  We also note that
above base temperature the RPA function agrees with the temperature dependence of the activated triplet-quintet excitation, $\omega_2$,
although shifted to higher energy transfers,
implying that these excitations may share a similar temperature
dependent dispersion renormalization.

\section{CONCLUSIONS}
Through INS measurements we have shown that there exists a well-defined singlet-triplet
spectrum in Ba$_3$Mn$_2$O$_8$.  Although the measured bandwidth of the magnetic spectrum is larger than the spin-gap, the exchange constants
indicate that Ba$_3$Mn$_2$O$_8$ can be considered a
triangular lattice of
weakly coupled $S=1$ dimers.
Comparison to an appropriate
RPA coupled dimer description of the scattering function is able
to determine the exchange constants.
In addition, the wave vector dependent scattering
intensity agrees with the dominant dimer bond being the
predicted short
vertical bond illustrated in Fig.~\ref{fig:structure}(b).

The examination of temperature dependent scattering intensity indicates
that both singlet-triplet and triplet-quintet excitations are
observed in the INS spectrum (there is no INS cross-section
for $S=1$ antiferromagnetic singlet-quintet excitations).
Based upon the relative energy scales of the two observed excitations,
we can estimate the mean energy of singlet-quintet
excitations as $\approx 3.4(4)$~meV.
Single-crystal INS measurements may be able to determine
the dispersion associated with
thermally activated triplet-quintet excitations or
perhaps observe multi-particle excitations,
quintet-triplet decay or interference of single-
and multi-particle excitations \cite{stonenaturephcc}.
The existence of
singlet-quintet and triplet-quintet excitations may also be
able to explain the heat capacity above
$T\approx 3$~K~which can not be accounted for by single excitation
models\cite{Tsujiiprb2005}.

We also point out that the
currently determined exchange constants may be
able to further describe the magnetic field dependent phase
diagram at low-temperatures or place limits
on the nature of the proposed long-range-ordered
phases.  The difference between the observed phase diagrams for $H \| c$
and $H \perp c$ must be related to the single ion anisotropy term in
the Hamiltonian, Eq.~\ref{eq:hamiltonian}. As we will show elsewhere\cite{samulonpaper}, this term
induces an effective exchange anisotropy in the low-energy Hamiltonian
that results from projecting the original Hamiltonian into the subspace
generated by the singlet and $S_z = 1$ triplet of each dimer.  These two
states can be described with a pseudospin 1/2 variable.
The combined effect of geometric frustration and
anisotropy leads to the appearance of a new phase for $H\perp c$ that
will be discussed in Ref.~\onlinecite{samulonpaper}.

\section{Acknowledgments}
MBS and MDL acknowledge valuable discussions with I. Zaliznyak and A. Zheludev.  ORNL is managed
for the US DOE by UT-Battelle Inc. under contract DE-AC05-00OR22725.
This work utilized facilities supported in part by the National
Science Foundation under Agreement No. DMR-0454672.  work at Stanford was supported by the National Science Foundation, under grant DMR 0705087.

\end{document}